\documentclass[12pt]{article}
\newcommand{\nc}{\newcommand}
\nc{\renc}{\renewcommand}

\nc{\half}{{\textstyle{1\over2}}}
\nc{\etal}{\mbox{\it et al. }}
\nc{\ie}{{\it i.e.}}
\nc{\eg}{{\it e.g.}}

\renc{\thefootnote}{\arabic{footnote}}
\nc{\capt}[1]{{\bf Figure.} {\small\sl #1}}

\nc{\eqs}[2]{\mbox{Eqs.~(\ref{#1},\,\ref{#2})}}
\nc{\eq}[1]{\mbox{Eq.~(\ref{#1})}}

\nc{\figs}[2]{\mbox{Figs.~(\ref{#1},\,\ref{#2})}}
\nc{\fig}[1]{\mbox{Fig~.(\ref{#1})}}

\nc{\tag}[1]{\label{#1} \marginpar{{\footnotesize #1}}}
\nc{\mtag}[1]{\label{#1} \mbox{\marginpar{{\footnotesize #1}}}}
\renc{\baselinestretch}{1.5}
\newlength{\overeqskip}
\newlength{\undereqskip}
\nc{\be}[1]{\begin{equation} \mbox{$\label{#1}$}}
\nc{\bea}[1]{\begin{eqnarray} \mbox{$\label{#1}$}}
\nc{\Section}[2]{\section{#2}\label{#1}}
\nc{\Bibitem}[1]{\bibitem{#1}}
\nc{\Label}[1]{\label{#1}}

\nc{\eea}{\vspace{\undereqskip}\end{eqnarray}}
\nc{\ee}{\vspace{\undereqskip}\end{equation}}
\nc{\bdm}{\begin{displaymath}}
\nc{\edm}{\end{displaymath}}
\nc{\dpsty}{\displaystyle}
\nc{\bc}{\begin{center}}
\nc{\ec}{\end{center}}
\nc{\ba}{\begin{array}}
\nc{\ea}{\end{array}}
\nc{\bab}{\begin{abstract}}
\nc{\eab}{\end{abstract}}
\nc{\btab}{\begin{tabular}}
\nc{\etab}{\end{tabular}}
\nc{\bit}{\begin{itemize}}
\nc{\eit}{\end{itemize}}
\nc{\ben}{\begin{enumerate}}
\nc{\een}{\end{enumerate}}
\nc{\bfig}{\begin{figure}}
\nc{\efig}{\end{figure}}
\nc{\arreq}{&\!=\!&}
\nc{\arrmi}{&\!-\!&}
\nc{\arrpl}{&\!+\!&}
\nc{\arrap}{&\!\!\!\approx\!\!\!&}
\nc{\non}{\nonumber\\*}
\nc{\align}{\!\!\!\!\!\!\!\!&&}

\def\lsim{\; \raise0.3ex\hbox{$<$\kern-0.75em
      \raise-1.1ex\hbox{$\sim$}}\; }
\def\gsim{\; \raise0.3ex\hbox{$>$\kern-0.75em
      \raise-1.1ex\hbox{$\sim$}}\; }
\nc{\DOT}{\hspace{-0.08in}{\bf .}\hspace{0.1in}}
\nc{\Laada}{\hbox {$\sqcap$ \kern -1em $\sqcup$}}
\nc\loota{{\scriptstyle\sqcap\kern-0.55em\hbox{$\scriptstyle\sqcup$}}}
\nc\Loota{{\sqcap\kern-0.65em\hbox{$\sqcup$}}}
\nc\laada{\Loota}
\nc{\qed}{\hskip 3em \hbox{\BOX} \vskip 2ex}

\nc{\real}{{\rm I \! R}}
\nc{\Z}{{\sf Z \!\!\! Z}}
\nc{\complex}{{\rm C\!\!\! {\sf I}\,\,}}
\def\bigid{\leavevmode\hbox{\small1\kern-3.8pt\normalsize1}}
\def\id{\leavevmode\hbox{\small1\kern-3.3pt\normalsize1}}
\nc{\slask}{\!\!\!/}
\nc{\bis}{{\prime\prime}}
\nc{\pa}{\partial}
\nc{\na}{\nabla}
\nc{\ra}{\rangle}
\nc{\la}{\langle}
\nc{\goto}{\rightarrow}
\nc{\swap}{\leftrightarrow}

\nc{\EE}[1]{ \mbox{$\cdot10^{#1}$} }
\nc{\abs}[1]{\left|#1\right|}
\nc{\at}[2]{\left.#1\right|_{#2}}
\nc{\norm}[1]{\|#1\|}
\nc{\abscut}[2]{\Abs{#1}_{\scriptscriptstyle#2}}
\nc{\vek}[1]{{\rm\bf #1}}
\nc{\integral}[2]{\int\limits_{#1}^{#2}}
\nc{\inv}[1]{\frac{1}{#1}}
\nc{\dd}[2]{{{\partial #1}\over{\partial #2}}}
\nc{\ddd}[2]{{{{\partial}^2 #1}\over{\partial {#2}^2}}}
\nc{\dddd}[3]{{{{\partial}^2 #1}\over
        {\partial #2 \partial #3}}}
\nc{\dder}[2]{{{d #1}\over{d #2}}}
\nc{\ddder}[2]{{{d^2 #1}\over{d {#2}^2}}}
\nc{\dddder}[3]{{d^2 #1}\over
        {d #2 d #3}}
\nc{\dx}[1]{d\,^{#1}x}
\nc{\dy}[1]{d\,^{#1}y}
\nc{\dz}[1]{d\,^{#1}z}
\nc{\dl}[1]{\frac{d\,^{#1}l}{(2\pi)^{#1}}}
\nc{\dk}[1]{\frac{d\,^{#1}k}{(2\pi)^{#1}}}
\nc{\dq}[1]{\frac{d\,^{#1}q}{(2\pi)^{#1}}}

\nc{\cc}{\mbox{$c.c.$ }}
\nc{\hc}{\mbox{$h.c.$ }}
\nc{\cf}{cf.\ }
\nc{\erfc}{{\rm erfc}}
\nc{\Tr}{{\rm Tr\,}}
\nc{\tr}{{\rm tr\,}}
\nc{\pol}{{\rm pol}}
\nc{\sign}{{\rm sign}}
\nc{\bfT}{{\bf T }}

\def\GeV{{\rm\ GeV}}

\def\TeV{{\rm\ TeV}}

\nc{\cA}{{\cal A}}
\nc{\cB}{{\cal B}}
\nc{\cD}{{\cal D}}
\nc{\cE}{{\cal E}}
\nc{\cG}{{\cal G}}
\nc{\cH}{{\cal H}}
\nc{\cL}{{\cal L}}
\nc{\cO}{{\cal O}}
\nc{\cT}{{\cal T}}
\nc{\cN}{{\cal N}}
\nc{\rvac}[1]{|{\cal O}#1\rangle}
\nc{\lvac}[1]{\langle{\cal O}#1|}
\nc{\rvacb}[1]{|{\cal O}_\beta #1\rangle}
\nc{\lvacb}[1]{\langle{\cal O}_\beta #1 |}
\nc{\bb}{\bar{\beta}}
\nc{\bt}{\tilde{\beta}}
\nc{\ctH}{\tilde{\cal H}}
\nc{\chH}{\hat{\cal H}}
\nc{\1}{\aa}
\nc{\2}{\"{a}}
\nc{\3}{\"{o}}
\nc{\4}{\AA}
\nc{\5}{\"{A}}
\nc{\6}{\"{O}}
\nc{\al}{\alpha}
\nc{\g}{\gamma}
\nc{\Del}{\Delta}
\nc{\e}{\epsilon}
\nc{\eps}{\epsilon}
\nc{\lam}{\lambda}
\nc{\om}{\omega}
\nc{\Om}{\Omega}
\nc{\ve}{\varepsilon}
\nc{\mn}{{\mu\nu}}
\nc{\vp}{\varphi}

\nc{\advp}[3]{{\it  Adv.\ in\ Phys.\ }{{\bf #1} {(#2)} {#3}}}
\nc{\annp}[3]{{\it  Ann.\ Phys.\ (N.Y.)\ }{{\bf #1} {(#2)} {#3}}}
\nc{\apl}[3]{{\it  Appl. Phys. Lett. }{{\bf #1} {(#2)} {#3}}}
\nc{\apj}[3]{{\it  Ap.\ J.\ }{{\bf #1} {(#2)} {#3}}}
\nc{\apjl}[3]{{\it  Ap.\ J.\ Lett.\ }{{\bf #1} {(#2)} {#3}}}
\nc{\app}[3]{{\it Astropart.\ Phys.\ }{{\bf #1} {(#2)} {#3}}}
\nc{\cmp}[3]{{\it  Comm.\ Math.\ Phys.\ }{{ \bf #1} {(#2)} {#3}}}
\nc{\cqg}[3]{{\it  Class.\ Quant.\ Grav.\ }{{\bf #1} {(#2)} {#3}}}
\nc{\epl}[3]{{\it  Europhys.\ Lett.\ }{{\bf #1} {(#2)} {#3}}}
\nc{\ijmp}[3]{{\it Int.\ J.\ Mod.\ Phys.\ }{{\bf #1} {(#2)} {#3}}}
\nc{\ijtp}[3]{{\it Int.\ J.\ Theor.\ Phys.\ }{{\bf #1} {(#2)} {#3}}}
\nc{\jmp}[3]{{\it  J.\ Math.\ Phys.\ }{{ \bf #1} {(#2)} {#3}}}
\nc{\jpa}[3]{{\it  J.\ Phys.\ A\ }{{\bf #1} {(#2)} {#3}}}
\nc{\jpc}[3]{{\it  J.\ Phys.\ C\ }{{\bf #1} {(#2)} {#3}}}
\nc{\jap}[3]{{\it J.\ Appl.\ Phys.\ }{{\bf #1} {(#2)} {#3}}}
\nc{\jpsj}[3]{{\it J.\ Phys.\ Soc.\ Japan\ }{{\bf #1} {(#2)} {#3}}}
\nc{\lmp}[3]{{\it Lett.\ Math.\ Phys.\ }{{\bf #1} {(#2)} {#3}}}
\nc{\mpl}[3]{{\it  Mod.\ Phys.\ Lett.\ }{{\bf #1} {(#2)} {#3}}}
\nc{\ncim}[3]{{\it  Nuov.\ Cim.\ }{{\bf #1} {(#2)} {#3}}}
\nc{\np}[3]{{\it  Nucl.\ Phys.\ }{{\bf #1} {(#2)} {#3}}}
\nc{\npps}[3]{{\it  Nucl.\ Phys.\ Proc.\ Suppl.\ }{{\bf #1} {(#2)} {#3}}}
\nc{\pr}[3]{{\it Phys.\ Rev.\ }{{\bf #1} {(#2)} {#3}}}
\nc{\pra}[3]{{\it  Phys.\ Rev.\ A\ }{{\bf #1} {(#2)} {#3}}}
\nc{\prb}[3]{{\it  Phys.\ Rev.\ B\ }{{{\bf #1} {(#2)} {#3}}}}
\nc{\prc}[3]{{\it  Phys.\ Rev.\ C\ }{{\bf #1} {(#2)} {#3}}}
\nc{\prd}[3]{{\it  Phys.\ Rev.\ D\ }{{\bf #1} {(#2)} {#3}}}
\nc{\prl}[3]{{\it Phys.\ Rev.\ Lett.\ }{{\bf #1} {(#2)} {#3}}}
\nc{\pl}[3]{{\it  Phys.\ Lett.\ }{{\bf #1} {(#2)} {#3}}}
\nc{\prep}[3]{{\it Phys.\ Rep.\ }{{\bf #1} {(#2)} {#3}}}
\nc{\prsl}[3]{{\it Proc.\ R.\ Soc.\ London\ }{{\bf #1} {(#2)} {#3}}}
\nc{\ptp}[3]{{\it  Prog.\ Theor.\ Phys.\ }{{\bf #1} {(#2)} {#3}}}
\nc{\ptps}[3]{{\it  Prog\ Theor.\ Phys.\ suppl.\ }{{\bf #1} {(#2)} {#3}}}
\nc{\physa}[3]{{\it  Physica\ A\ }{{\bf #1} {(#2)} {#3}}}
\nc{\physb}[3]{{\it  Physica\ B\ }{{\bf #1} {(#2)} {#3}}}
\nc{\phys}[3]{{\it Physica\ }{{\bf #1} {(#2)} {#3}}}
\nc{\rmp}[3]{{\it  Rev.\ Mod.\ Phys.\ }{{\bf #1} {(#2)} {#3}}}
\nc{\rpp}[3]{{\it Rep.\ Prog.\ Phys.\ }{{\bf #1} {(#2)} {#3}}}
\nc{\sjnp}[3]{{\it Sov.\ J.\ Nucl.\ Phys.\ }{{\bf #1} {(#2)} {#3}}}
\nc{\spjetp}[3]{{\it Sov.\ Phys.\ JETP\ }{{\bf #1} {(#2)} {#3}}}
\nc{\yf}[3]{{\it Yad.\ Fiz.\ }{{\bf #1} {(#2)} {#3}}}
\nc{\zetp}[3]{{\it Zh.\ Eksp.\ Teor.\ Fiz.\  }{{\bf #1}  {(#2)} {#3}}}
\nc{\zp}[3]{{\it Z.\ Phys.\ }{{\bf #1} {(#2)} {#3}}}
\nc{\ibid}[3]{{\sl ibid.\ }{{\bf #1} {#2} {#3}}}
\nc{\rf}[1]{(\ref{#1})}
\nc{\nn}{\nonumber \\*}
\nc{\bfB}{\bf{B}}
\nc{\bfv}{\bf{v}}
\nc{\bfx}{\bf{x}}
\nc{\bfy}{\bf{y}}
\nc{\vx}{\vec{x}}
\nc{\vy}{\vec{y}}
\nc{\oB}{\overline{B}}
\nc{\oI}{\overline{I}}
\nc{\oR}{\overline{R}}
\nc{\rar}{\rightarrow}
\nc{\ti}{\times}
\nc{\slsh}{\hskip-5pt/}
\nc{\sm}{Standard~Model~}
\nc{\MP}{M_{\rm Pl}}
\nc{\tp}{t_{\rm Pl}}
\nc{\ave}{\bar{E}}

\nc{\eff}{{\rm eff}}
\nc{\kk}{\vek{k}}
\nc{\pp}{{\rm p}}
\nc{\ga}{g_{a\gamma}}
\nc{\vv}{\\}
\nc{\eee}{{\bf E}}
\nc{\bbb}{{\bf B}}
\nc{\qcd}{T_{\rm QCD}}
\nc{\G}{\rm \ G}
\def\vec#1{{\bf #1}}

\def\lae{\;^{<}_{\sim} \;} \def\gae{\; ^{>}_{\sim} \;} 

\def\ell{e^{c}LL}

\begin{document}
{\title{\vskip-2truecm{\hfill {{\small \\
	\hfill \\
	}}\vskip 1truecm}
{\LARGE Right-Handed Sneutrino Condensate Cold Dark Matter and the Baryon-to-Dark Matter Ratio.
}     
}}
{\author{
{\sc  John McDonald$^{1}$}\\
{\sl\small Cosmology and Astroparticle Physics Group, University of Lancaster,
Lancaster LA1 4YB, UK}
}
\maketitle
\begin{abstract}
\noindent

            The similarity of the observed densities of baryons and cold dark matter suggests that they have a common or related origin. This can be understood in the context of the MSSM with right-handed (RH) sneutrinos if cold dark matter is due to a $d = 4$ flat direction condensate of very weakly coupled RH sneutrino LSPs and the baryon asymmetry is generated by Affleck-Dine leptogenesis along the $d = 4$ $\left(H_{u}L\right)^{2}$ flat direction. The correct density of RH sneutrino dark matter is obtained if the reheating temperature is in the range $10^{6}-10^{8} \GeV$. A cold dark matter isocurvature perturbation close to present observational bounds is likely in the case of inflation driven by a D-term or by an F-term with suppressed Hubble corrections to the A-terms. An observable baryon isocurvature perturbation is also possible in the case of D-term inflation models.

\end{abstract} 
\vfil
 \footnoterule {\small $^1$j.mcdonald@lancaster.ac.uk}   
 \newpage 
\setcounter{page}{1}

\section{Introduction}  

      A striking feature of the observed Universe is the similar mass density in baryons and cold dark matter (CDM). From the WMAP three-year results for the $\Lambda$CDM model, 
 $\Omega_{DM}/\Omega_{B} = 5.65 \pm 0.58$ \cite{bdm}.  In conventional models of baryogenesis and dark matter the physical mechanisms behind the respective densities are typically unrelated \cite{ktb}. For example, this is generally true in the
case where dark matter originates from freeze-out of a thermal equilibrium density of weakly-interacting particles. As there is no reason to expect the values of $\Omega_{B}$ and $\Omega_{DM}$ from entirely unrelated physical processes to be within an order of magnitude of each other, there is an implicit acceptance of either a remarkable coincidence or an undefined anthropic selection mechanism. 

      A more obvious interpretation of the baryon-to-dark matter ratio is that the densities originate via a common physical 
mechanism. Such an interpretation could provide a strong principle by which to identify both a viable particle physics model and the explanation of dark matter and baryogenesis within that model. Should a given particle physics model naturally have within its structure a common mechanism for the origin of dark matter and baryogenesis, then both the model and the associated mechanism would be strongly favoured. 

           In this paper we will apply this principle to the minimal supersymmetric (SUSY) Standard Model (MSSM) \cite{nilles} to determine the most likely cold dark matter particle and mechanism for baryogenesis. We will consider the MSSM extended by right-handed (RH) neutrino superfields in order to generate neutrino masses ($\nu$MSSM).

                     It has recently been proposed that the lightest supersymmetric particle (LSP) could be a RH sneutrino \cite{moroi}.  The scenario discussed in \cite{moroi} considered the case where the neutrino masses are of Dirac-type, with no SUSY mass term for the RH neutrino superfield. Since the RH sneutrino mass is purely due to soft SUSY breaking, the RH sneutrino can then be the LSP.  This model may have striking consequences for collider phenomenology: the next-to-lightest SUSY particle of the MSSM sector (MSSM-LSP) would have collider signatures identical to a conventional LSP, but could be a coloured or charged particle instead of a neutralino \cite{moroi}. 

               The dark matter RH sneutrinos considered in \cite{moroi} originated from the decay of thermal equilibrium MSSM particles, although the possibility of dark matter being due to a RH sneutrino condensate
was also noted. In the following we will argue that the baryon-to-dark matter ratio favours a RH sneutrino condensate as the explanation of cold dark matter.  This will be true if there exists a mechanism within the $\nu$MSSM by which a baryon density can be generated which is naturally of the same magnitude as the RH sneutrino dark matter density. We will show that this can be achieved via Affleck-Dine (AD) leptogenesis \cite{ad} along the $(H_{u}L)^{2}$ flat direction \cite{drt,fd}. Large CDM and/or baryonic isocurvature perturbations are possible in this framework, depending of the structure of the inflation model.

                     There exist other models which attempt to explain the baryon-to-dark matter ratio. 
The possibility that sneutrinos can play the role of dark matter has been considered in \cite{west,ap}. In 
\cite{west} a mixed sneutrino dark matter particle with a large RH sneutrino component was considered. In this model thermal relic densities of sneutrinos and baryons of the right magnitude can be generated   
from an initial lepton asymmetry. This requires that the weak scale has the right magnitude in order to annihilate the sneutrino asymmetry sufficiently. In \cite{ap} a RH sneutrino condensate was considered in a lepton-number conserving renormalizable extension of the MSSM. This model shares the features of Affleck-Dine leptogenesis and RH sneutrino condensates with the model proposed here, but employs a different AD leptogenesis mechanism and is dynamically unrelated. In this model the number density of dark matter particles is related to the number density of baryons via lepton number conservation. This type of relation, based on a conserved charge, requires a very light LSP of a few GeV in mass, which is disfavoured in gravity-mediated SUSY breaking models. In \cite{ke} the possibility that the late-decaying Q-balls which form along $d=6$ flat directions of the MSSM during AD baryogenesis could produce dark matter neutralinos was considered. This model also relates the number density of baryons and neutralinos and as a result requires a neutralino mass $\approx 2 \GeV$, which is experimentally ruled out in the context of the MSSM. Recently a solution to this problem was proposed where naturally light axinos from the decay of MSSM Q-balls could account for the baryon-to-dark matter ratio \cite{lesosam}. An earlier SUSY model is based on the CP-violating decay of a condensate of massive scalar particles to baryons and dark matter \cite{thomas}.  Non-SUSY models also exist. 
In \cite{tytgat}, a model based on the universal see-saw mechanism was proposed in which the baryon number is related 
to the number of dark matter particles via baryon number conservation and the decay of TeV-scale quarks to 
right-handed Majorana neutrinos of mass $\approx 1 \GeV$. A similar mechanism was proposed in \cite{klow}. In \cite{farrar} it was proposed that the dark matter particle number density could be related to the baryon number density 
if the dark matter particles carry baryon number and the annihilation cross-sections for dark matter baryons and anti-baryons differ. Eariler models include those based on electroweak baryogenesis \cite{kaplan} and anomalous $(B+L)$-violation \cite{barr}.

     A striking feature of all of these models (with the exception of \cite{west}) is that they strictly relate the number of baryon to the number of dark matter particles via a conserved charge. In the case of SUSY models this results in unfavourably small LSP masses. In contrast, the model we present here is not based on a conserved charge but instead is purely dynamical in nature, based on the similar dynamics of the seperate flat direction fields responsible for the baryon asymmetry and dark matter density. Information from accelerators, in particular evidence of SUSY and the nature of the LSP from the Large Hadron Collider (LHC), and astronomical observations, such as evidence of isocurvature perturbations, could in principle distinguish between the possible models.

           The paper is organised as follows. In Section 2 we discuss dark matter from a RH sneutrino condensate and AD leptogenesis from the $(H_{u}L)^{2}$ flat direction. We calculate the baryon-to-dark matter ratio and the range of parameters for which the densities of dark matter and baryons are similar. In Section 3 we calculate the magnitude of the CDM and baryon isocurvature perturbations expected in the model. In Section 4 we present our conclusions.

\section{Cold Dark Matter and the Baryon Asymmetry from $d=4$ Flat Directions}

          The key feature of the $\nu$MSSM is that both the baryon asymmetry and cold dark matter can originate from condensates of scalar fields along flat directions of the scalar potential. In particular, if the flat directions responsible are both lifted by non-renormalisable superpotential terms of the same dimension, then we will show that the baryon and dark matter densities are automatically related.

                Since the RH sneutrino is a gauge singlet, it is  natural to consider a non-renormalisable superpotential term of the lowest possible dimension, $d = 4$. In order to account for the baryon-to-dark matter ratio, the baryon asymmetry must then originate via AD leptogenesis along the $d=4$ $(H_{u}L)^{2}$ flat direction\footnote{In order to have an $(H_{u}L)^{2}$ flat direction in the presence of a large amplitude for $N$ in the $\nu$MSSM, the $N$ field would have consist of combination of $N$ generations that does not couple to the flat direction $L$ via $\lambda_{\nu}$. However, since we are considering very small values of $\lambda_{\nu}$ in the following, $\lambda_{\nu} \lae 10^{-12}$, in practice any lifting of the $(H_{u}L)^{2}$ flat direction by the $N$ amplitude is negligible.}. The superpotential we consider is therefore
\be{e1}  W = W_{\nu MSSM} + W_{NR}      ~,\ee
where $W_{\nu MSSM}$ is the $\nu$MSSM superpotential 
\be{e2}  W_{\nu MSSM} = \lambda_{\nu} NH_{u}L + \lambda_{e}e^{c}H_{d}L + \lambda_{u}u^{c}H_{u}Q + \lambda_{d}d^{c}H_{d}Q   
+ \mu H_{u}H_{d} + \frac{M_{N}}{2} N^{2}     ~\ee 
and $W_{NR}$ contains Planck-suppressed non-renormalisable terms\footnote{We include $1/4!$ factors so that the 
physical strength of the interactions is dimensionally determined by $M$.}  
\be{e3} W_{NR} = \frac{\lambda_{N}}{4!} \frac{N^{4}}{M} + \frac{\lambda_{\Phi}}{4 !}  \frac{\Phi^{4}}{M}      ~.\ee 
Here $N$ is the RH neutrino superfield, $\Phi$ is the flat direction AD superfield
of the $d=4$ $(H_{u}L)^{2}$ flat direction and $M = M_{Pl}/\sqrt{8 \pi} \approx 2.4 \times 10^{18} \GeV$. For simplicity we have suppressed generation indices. $N$ carries lepton number $L = 1$. This is the most general renormalisable superpotential which is invariant under standard R-parity, which we assume is unbroken in order to have a stable LSP.

     $M_{N}$ cannot be larger than the 
soft SUSY breaking mass scale, $m_{s} \approx 100 \GeV - 1 \TeV$, if the RH sneutrino is to be the LSP.  The simplest possibility is that there is no $N^{2}$ term in the superpotential, in which case the neutrinos gain Dirac masses via very small Yukawa couplings, $\lambda_{\nu} \lae 10^{-12}$. This can be understood if the renormalisable part of the superpotential has an R-symmetry under which $R(u^{c},d^{c},e^{c},Q,L,N) = 1/2$, $R(H_{u},H_{d}) = 1$. The R-symmetry is assumed to be broken by the Planck-suppressed non-renormalisable terms. Alternatively, lepton number may be conserved in the renormalisable superpotential but broken by the non-renormalisable terms. In order to generate a baryon density similar to the dark matter density we need a $d=4$ $B-L$-violating flat direction. For the conventional MSSM flat directions (those independent of $N$), the only $B-L$-violating $d=4$ monomial is $(H_{u}L)^{2}$ \cite{drt,fd}. In addition there are $B-L$ conserving MSSM flat directions corresponding to the monomials $QQQL$ and $u^{c}u^{c}d^{c}e^{c}$. However, unless there is a suppression of these superpotential terms for 1st and 2nd generation superfields they will induce too rapid proton decay \cite{pdecay,pdecay2}. The $QQQL$ and $u^{c}u^{c}d^{c}e^{c}$ terms can be eliminated by an anomaly-free discrete gauge symmetry of the MSSM (which is left unbroken by quantum gravity effects) which permits the $(H_{U}L)^{2}$ operator 
\cite{pdecay2,luhn}. Under the requirements that baryon-number is conserved up to dimension 5 operators and that the MSSM $\mu$-term and $(H_{U}L)^{2}$ operator are permitted, the possible anomaly-free discrete gauge symmetries are baryon triality \cite{pdecay2}, which requires seperate imposition of R-parity (also anomaly-free) to obtain the MSSM, or baryon hexality, which automatically reproduces the low-energy R-parity conserving superpotential \cite{luhn}.

\subsection{Scalar Potential for $d=4$ Flat Direction} 

For a $d = 4$ flat direction lifted by a superpotential of the form,  
\be{e4a}  W_{\Psi} = \frac{\lambda_{\Psi}}{4!} \frac{\Psi^{4}}{M}     ~,\ee
the generic scalar potential with gravity-mediated soft SUSY-breaking terms and order $H$ corrections from supergravity \cite{drt,h2a,h2b} is of the form 
\be{e4} V(\Psi) = \left( m_{\Psi}^{2} - c_{\Psi}H^{2}\right) 
\left|\Psi\right|^{2} + \left( A_{\Psi}\frac{ \lambda_{\Psi}}{4!M} \Psi^{4} + h.c. \right) + \frac{ \left|\lambda_{\Psi} \right|^{2}}{3!^{2} M^{2}} \left|\Psi\right|^{6}           ~,\ee
where $\Psi$ ($ \equiv N$ or $\Phi$) denotes the flat direction superfield of interest. The couplings $c_{\Psi}$ and $\lambda_{\Psi}$
are usually assumed to have magnitude of order 1, although smaller values of $\lambda_{\Psi}$ could conceivably occur depending on the Planck-scale physics responsible for the non-renormalisable terms. The form of the order $H$ corrections during and after inflation will depend on the origin of the energy density driving inflation (i.e. whether it is driven by an F- or D-term in the scalar potential\footnote{In this paper F- and D-term inflation will refer to the term in the inflaton scalar potential responsible for driving inflation. F-term and D-term hybrid inflation models are specific examples of these.}) and the K\"ahler potential. In D-term inflation $|c_{\Psi}| = 0$ during inflation and $|c_{\Psi}| \approx 1$ after inflation. For F-term inflation $|c_{\Psi}| \approx 1$ during and after inflation \footnote{F-term inflation models can have $|c_{\Psi}| = 0$ during and after inflation if there is a Heisenberg symmetry of the K\"ahler potential \cite{oliveh}}. In order not to suppress the amplitude of the RH sneutrino or Affleck-Dine scalar we will consider the case where $c_{\Psi}$ is positive. In this case the scalar potential has a minimum after inflation ends. For $H > m_{\Psi}$ the minimum of the potential is given by
\be{e5} |\Psi|_{min}  \approx \left( \frac{12 c_{\Psi}}{ \lambda_{\Psi}^{2}} \right)^{1/4} \left(H M\right)^{1/2}            ~.\ee
The $\Psi \neq 0$ minimum of the potential will vanish at $H = H_{osc \; \Psi} \approx m_{\Psi}/c_{\Psi}^{1/2}$, at which time coherent oscillations of the $\Psi$ field begin. The initial amplitude of the coherently oscillating $\Psi$ field is then   
\be{e5a}    |\Psi|_{osc} \approx   |\Psi|_{min} [H \approx H_{osc \;\Psi}] = \left(\frac{12}{\lambda_{\Psi}^{2}}\right)^{1/4} \left(m_{\Psi}M\right)^{1/2} 
~.\ee
The initial energy density in the oscillating field is then  
\be{e5b}    \rho_{\Psi\;osc} = m_{\Psi}^{2} |\Psi|_{osc}^{2} \approx \frac{\sqrt{12}}{\lambda_{\Psi}} m_{\Psi}^{3} M      ~.\ee

\subsection{CDM from a $d=4$ RH Sneutrino Condensate}

     The RH sneutrino is the unique candidate for condensate dark matter in the $\nu$MSSM.  Conventional MSSM flat directions are linear combinations of squark and slepton fields, which are generally excluded as dark matter candidiates \cite{dmex}. In contrast, the RH sneutrino can have very weak Yukawa couplings and is a gauge-singlet. It can therefore easily evade all direct constraints on dark matter particles. In particular, the model with $M_{N} = 0$ has Dirac neutrino masses and therefore extremely small neutrino Yukawa couplings, $\lambda_{\nu} \lae 10^{-12}$,  making the RH sneutrino an excellent candidate for condensate dark matter in this case.  

      Coherent oscillations of the RH sneutrino begin before radiation domination if the reheating temperature satisfies the gravitino bound $T_{R} \lae 10^{6-8} \GeV$ \cite{morgrav}, since in this case $H(T_{R}) \lae 0.1 \GeV \ll m_{N}$. The present energy density in the RH sneutrino condensate is therefore
\be{e6}    \rho_{N\;o} = \left( \frac{a_{osc}}{a_{R}} \right)^{3}
\left( \frac{a_{R}}{a_{o}} \right)^{3}  \rho_{N\;osc}  
= \frac{c_{N} k_{T_{\gamma}}^{2} T_{\gamma}^{3} T_{R} }{M^{2} m_{N}^{2}} \rho_{N\;osc}           ~,\ee 
where $a_{R}$ is the scale factor at radiation-domination, $a_{o}$ is the scale factor at present, $T_{\gamma}$ is the present photon temperature, 
$\rho_{c} =  8.1 \times h^{2} 10^{-47} GeV^{4}$ is the critical density for a flat Universe, $\rho_{N \; osc} = m_{N}^{2} |N|_{osc}^{2}$ and $k_{T} = 
\left( \pi^{2}g\left(T\right)/90\right)^{1/2}$ with $g(T)$ is the number of degrees of freedom in thermal equilibrium.  The reheating temperature required to have a RH sneutrino condensate density $\Omega_{N}$ is therefore 
\be{e7}   T_{R} \approx \left(\frac{\lambda_{N}^{2}}{12}\right)^{1/2} \frac{ M  \Omega_{N} \rho_{c}}{c_{N} k_{T_{\gamma}}^{2} T_{\gamma}^{3} m_{N}}   ~.\ee  
With typical values for the parameters this becomes
\be{e7a}  T_{R} \approx 2.6 \times 10^{7} \; \frac{\lambda_{N}}{c_{N}} \left(\frac{h}{0.7}\right)^{2} \left(\frac{\Omega_{N}}{0.23}\right) \left(\frac{100 \GeV}{m_{N}}\right) \GeV    ~.\ee  
For example, if $\lambda_{N}/c_{N}$ is in the range 0.1 to 1 then a value of $T_{R}$ in the range $10^{6}-10^{8} \GeV$ is necessary to account for dark matter. This range of $T_{R}$ is compatible with the gravitino upper bound on the reheating temperature, even for the case where the gravitino decays primarily to hadrons \cite{morgrav}.

\subsection{$d = 4$ $(H_{u}L)^{2}$ Affleck-Dine Leptogenesis} 
    
       The A-term proportional to $\Phi^{4} \sim (H_{u}L)^{2}$ breaks $L$ and is explicitly dependent upon the phase of the $\Phi$ scalar. It can therefore induce a lepton asymmetry in the complex $\Phi$ condensate via the Affleck-Dine mechanism \cite{ad}. The contribution of the A-terms to the potential is comparable with the mass squared term just when the complex $\Phi$ condensate forms at $H = H_{osc \; \Phi} \approx m_{\Phi}/c_{\Phi}^{1/2}$. If we define the real direction of $\Phi$ to be that along which the gravity-mediated A-term is real and negative, 
then the CP violating phase responsible for the $L$ asymmetry approximately corresponds to the angle of the $\Phi$ field in the complex plane at $H = H_{osc \; \Phi}$ relative to this late-time real direction. The origin of this angle will differ depending on whether the A-term itself receives an order $H$ correction after the end of slow-roll inflation.

   The order $H$ correction to the A-term originates from cross-terms in the supergravity potential which couple the inflaton $S$ to the flat direction scalars $\Psi$. For $|S| \ll M$ the contribution of these terms can be expanded in powers of $S/M$ \cite{drt} 
\be{ex1} {\cal L} \sim \frac{1}{M} \int d^{4} \theta S \Psi^{\dagger}\Psi + O\left(\frac{1}{M^{2}}\right) \int d^{4} \theta S^{2} \Psi^{\dagger}\Psi   +   ...    ~.\ee 
This implies an A-term due to the inflaton F-term of the form 
\be{ex2}  A_{i} = A_{s \; i} + a_{i}\left(1 + O\left(\frac{S}{M}\right) + ...\right) H     ~,\ee
where $A_{s\;i}$ is the gravity-mediated A-term and $|a_{i}| \approx 1$. 
Therefore if there is a linear coupling of the inflaton superfield to the flat direction scalar in the effective theory 
then an order $H$ correction to the A-term will arise. However, if the linear term is suppressed then the leading A-term correction will be of order $(|S|H/M)$. In this case the A-term will have no dynamical effect if $|S|/M$ is sufficiently small during and after inflation, which we will show is the case if $|S|/M \lae 1/40$.  The suppression of order $H$ corrections to the A-terms can have significant consequences for isocurvature perturbations.   

      In the case with order $H$ corrections to the A-terms, the flat direction scalar will be at the minimum of the scalar potential along the direction which makes the $H$-dependent A-terms real and negative. So long as there is a complex phase between $A_{s \; i}$ and $a_{i}$, then at the onset of AD condensate formation (at which time the  $A_{s \; i}$ and $a_{i}H$ terms are comparable) the real direction will change from that defined by $a_{i}$ to that defined by $A_{s \; i}$ as $H$ decreases. Therefore initially the oscillating $\Phi$ field is not along the late-time real direction. Since at $H = H_{osc \; \Phi}$ the mass squared term and A-term in the scalar potential are of comparable magnitude, the A-term will cause the real and imaginary components of $\Phi$ to oscillate out of phase. This results in a condensate which describes an ellipse in the complex plane at late times, corresponding to an $L$ asymmetry in the AD condensate \cite{ad}.

    If there is a symmetry under which the inflaton transforms, such as a discrete symmetry or an $R$-symmetry, then the order $H$ corrrections to the A-term will be suppressed both during and after inflation. Such a symmetry is a common feature of SUSY inflation models e.g. an R-symmetry may protect the inflaton potential from non-renormalisable corrections in SUSY hybrid inflation models \cite{dti,fti,lr}. For sufficiently suppressed $H$ corrections to the A-terms, the complex phase producing the $L$-asymmetry will simply be the random initial phase of $\Phi$ which exists during inflation. 
  
               In either case, at $H \ll H_{osc \; \Phi}$ the $\Phi$ scalar will be of the form 
$\Phi(t) = \left(\phi_{1}(t) + i \phi_{2}(t) \right)/\sqrt{2}$, where 
\be{e8}                     \phi_{1} = \phi(t) \cos(\theta) \sin(m_{\Phi} t)  \;\; ; \;\;\;\; \phi_{2} = \phi(t) \sin(\theta) \sin(m_{\Phi} t + \delta)   \;\; ; \;\; \phi(t) \propto a^{-3/2}     ~.\ee 
Here $\theta$ is the initial phase of the $\Phi$ field relative to the late-time real direction and $\delta$ is a phase shift induced by the A-term; the precise value of $\theta$ and $\delta$ will be determined    
by the scalar field dynamics during the period when the $\Phi \neq 0$ minimum vanishes at $H \approx H_{osc\;\Phi}$.      
The resulting L asymmetry of the AD condensate is then       
\be{e9}   n_{L} = \frac{i}{2} \left( \dot{\Phi}^{\dagger} \Phi - \Phi^{\dagger} \dot{\Phi}  \right)
  = \frac{1}{4} m_{\Phi} \phi^{2}(t) \sin(2 \theta) \sin (\delta)  ~.\ee 
In this we have included an overall factor 1/2 since the $\Phi$ field carries lepton number $L = 1/2$. 
Once the AD condensate decays the lepton asymmetry will be converted into a baryon asymmetry by 
$B + L$ violating sphaleron processes, such that $n_{B} = n_{L\;{\rm initial}}/2$. The magnitude of the baryon asymmetry today is then  
\be{e10}    n_{B} =  \frac{f_{A}}{4} \frac{ \rho_{\Phi \; o} }{m_{\Phi}} \;\; ; \;\; f_{A} = \sin(2 \theta) \sin (\delta)     ~.\ee
In this we have defined $\rho_{\Phi \; o} = m_{\Phi}^{2} \phi(t_{o})^{2}/2$ to be the density that a coherently oscillating $\Phi$ field along the $d = 4$ flat direction would have at present if it did not decay, where $\phi(t_{o})$ is the amplitude the oscillating field at present. $f_{A}$ is a factor parameterising the CP-violation due to the A-term. For typical CP violating angles and initial conditions we expect $f_{A}$ to be in the range 0.1-1.

\subsection{Baryon-to-Dark Matter Ratio}

                         From \eq{e10} energy density in baryons at present is 
\be{e11}      \rho_{B\;o} = m_{n} n_{B} =  \frac{f_{A}}{4} \frac{m_{n}}{m_{\Phi}} \rho_{\Phi \; o}     ~,\ee
where $m_{n}$ is the nucleon mass. From this we can that the energy density in baryons is typically of the order of  
$m_{n}/m_{\Phi}$ times the energy density in a coherently oscillating $d=4$ flat direction scalar. Therefore 
if dark matter is due to a condensate along a $d = 4$ direction we will automatically have similar values of $\Omega_{DM}$ and $\Omega_{B}$. From \eq{e5b} and \eq{e6} we find that $\rho_{N\; o} \propto c_{N}m_{N}/\lambda_{N}$,  with an analogous expression for $\rho_{\Phi\;o}$. Therefore 
\be{e12}  \frac{\Omega_{B}}{\Omega_{DM}} = \frac{f_{A}}{4} \frac{m_{n}}{m_{\Phi}} 
\frac{\rho_{\Phi \; o}}{\rho_{N \; o}} = \frac{f_{A}}{4}\frac{m_{n}}{m_{N}} \frac{c_{\Phi}}{\lambda_{\Phi}} 
\frac{\lambda_{N}}{c_{N}} = \frac{f_{A}}{400} 
\left(\frac{100 \GeV}{m_{N}}\right) 
\frac{c_{\Phi}}{c_{N}}  \frac{\lambda_{N}}{\lambda_{\Phi}}    ~.\ee 
This shows that it is possible to explain the observed baryon-to-dark matter density ratio, $\Omega_{B}/\Omega_{DM} \approx 1/6$, with values of $\lambda_{N}$ and $\lambda_{\Phi}$ which are within an order of magnitude of each other.
For example, $\lambda_{N} \approx 10 \lambda_{\Phi}$, $c_{\Phi} \approx 6 c_{N}$, $f_{A} \approx 1$  
and $m_{N} \approx 100 \GeV$ gives $\Omega_{B}/\Omega_{DM} \approx 1/6$. If $\lambda_{\Phi}$ can be significantly less than $\lambda_{N}$, for example $\lambda_{\Phi} \approx 0.01\lambda_{N}$, then smaller values of $f_{A}$ and $c_{\Phi}$ are possible. 

    Thus the baryon-to-dark matter ratio can be accounted for if the couplings $\lambda$ for all $d=4$ flat directions of the $\nu$MSSM are within a few orders of magnitude of each other.    
A small hierarchy, with $\lambda_{\Phi}$ one to two orders of magnitude less than $\lambda_{N}$, is required. Since there could be a generational dependence of the non-renormalisable couplings, as in the case of renormalisable Yukawa couplings in the MSSM, it is reasonable to expect that $\lambda_{\Phi}$ and $\lambda_{N}$ could have significantly different values. 
Another possibility, which will be explored in future work, is that renormalization group evolution of the couplings of the $(H_{u}L)^{2}$ flat direction scalar and the RH sneutrino from the Planck scale to the scale of AD leptogenesis could result in the required hierarchy. 

         In the case with three generations of RH sneutrino, we expect condensates of all three RH sneutrino mass eigenstates to form. The 
heavier RH sneutrinos will decay to the LSP RH sneutrino. Any matrix element for the decay of the heavier RH sneutrinos to the LSP will be proportional to $\lambda_{\nu}^{2}$, therefore a very conservative upper bound on the decay rate is $\Gamma < \lambda_{\nu}^{4} m_{N}$. 
With $\lambda_{\nu} \approx 10^{-12}$ and $m_{N} \approx 100 \GeV$ this 
gives a lifetime $\tau > 10^{14}$ years. Therefore the heavier RH sneutrinos have a lifetime longer than the age of the Universe and so will also contribute to the dark matter density.

\section{Isocurvature Perturbations}

       Isocurvature perturbations of both the CDM and baryons can arise due to quantum fluctuations of the amplitude and phase of the flat direction scalar fields during inflation. The magnitude of the resulting isocurvature perturbation today will depend on the form of the order $H$ corrections to the mass squared terms and A-terms during and after inflation.

\subsection{CDM Isocurvature Perturbation}

       If all order $H$ corrections are zero during inflation, as in D-term inflation models, then the complex inflation field is massless and so there will be quantum fluctuations of both the phase and amplitude. 

           An alternative possibility is that the $H$ corrections to the A-terms are sufficiently suppressed both during and after inflation. In this case the quantum fluctuations of the phase of $N$ will be unsuppressed during and after inflation in both D-term and F-term inflation models. When the expansion rate drops to $H_{osc\;N}$, the A-term becomes dynamically important since the contribution of the A-term to the potential is of the same magnitude as the mass squared term at this time. The effect is that the quantum fluctuation in the phase field $\delta \theta$ will induce a fluctuation in the $N$ amplitude such that $\left|\delta N/N \right| = O(1) \delta \theta$.  (This mechanism was first discussed in the context of the phase-induced curvaton model \cite{pcurv}.) Therefore CDM isocurvature perturbations can also be generated in F-term inflation models.

        The condition for the phase fluctuations to be effectively unsuppressed is that the correlation length of the fluctuation at the end of inflation, $l \approx H^{-1} e^{3H^{2}/2 m^{2}} $, is larger than the region corresponding to the observed Universe, which requires that $3 H^{2}/2 m^{2} > 60$, where $m$ is the mass of the fluctuating field \cite{drt}. In the case of the phase field, an order $H$ A-term correction will produce an effective mass for the field at the minimum of the potential $m^{2} \approx H^{2}$. In the case of suppressed linear inflaton couplings, this effective mass becomes $m^{2} \approx |S| H^{2}/M$. Therefore the inflaton field during inflation must satisfy $|S|/M \lae 1/40$ for the A-term correction to be negligible.  

           To analyse the isocurvature perturbations we follow the discussion given in \cite{crotty}. Since the adiabatic perturbations come from inflaton quantum fluctuations while the CDM isocurvature perturbations come from unrelated quantum fluctuations of the RH sneutrino field, the adiabatic and CDM isocurvature perturbations will be uncorrelated. The angular power spectrum due to a mixture of uncorrelated adiabatic and CDM isocurvature perturbations is given by \cite{crotty} 
\be{e15} C_{l} = (1 - \alpha) C_{l}^{ad} + \alpha C_{l}^{iso}   ~,\ee
where 
\be{e15a}  \alpha = \frac{B^{2}}{1 + B^{2}}  \;\;\; ; \;\;\;     B^{2} = \frac{|{\cal S}(k)|^{2}}{|{\cal R}(k)|^{2}}                 ~.\ee
$C_{l}^{ad}$ and $C_{l}^{iso}$ are the multipole moments for the adiabatic and CDM isocurvature perturbation. $|{\cal S}(k)|^{2}$ and $|{\cal R}(k)|^{2}$ are the Fourier transforms of the mean squared entropy and curvature perturbation respectively. These are related to the power spectra by   
\be{e17}  |{\cal R}(k)|^{2} = \frac{2 \pi^{2} P_{{\cal R}}}{k^{3}} \;\;\; ; \;\; |{\cal S}(k)|^{2} = \frac{2 \pi^{2} 
P_{{\cal S}  }}{k^{3}}    ~.\ee
$P_{{\cal R}}$ is given by its COBE-normalised value, $P_{{\cal R}}^{1/2} = 4.8 \times 10^{-5}$ \cite{ll}.   

In the following we will consider the RH sneutrino dark matter to oscillate in the real direction, so that $N = N_{1} /\sqrt{2}$. For simplicity we will drop the subscript 1 and use $N$ to denote the canonically normalized real part of the complex field. 
The entropy perturbation is related to the 
fluctuation of the RH sneutrino field at the onset of coherent oscillations
\be{e19} {\cal S} = \left(\frac{\delta \rho}{\rho}\right)_{N\;iso} \equiv \frac{2 \delta N_{I}}{N_{I}}     ~.\ee
Here $N_{I}$ and $\delta N_{I}$ are the values of the amplitude and 
its fluctuation at horizon exit during inflation. In this we assume that $\delta N/N$ remains constant once the fluctuation exits the horizon, which we show below is true for $d=4$ directions.   
Therefore    
\be{e20}  P_{{\cal S} } \equiv P_{\left(\frac{2 \delta N_{I}}{N_{I}}\right) } =  \frac{4 P_{\delta N_{I}}}{N_{I}^{2}}  \;\;\; ; \;\;\;
 P_{\delta N_{I}} = \left( \frac{H_{I}}{2 \pi} \right)^{2}           ~, \ee
where $H_{I}$ is the value of $H$ during inflation, which for simplicity we assume to be constant. Therefore   
\be{e23}  |{\cal S}|^{2} = \frac{2 H_{I}^{2}}{k^{3} N_{I}^{2}}     ~.\ee 
and 
\be{e24}    B^{2} \equiv \frac{ |{\cal S}|^{2}  }{ |{\cal R}|^{2}  } = 
\frac{H_{I}^{2}}{\pi^{2} P_{\cal{R}} N_{I}^{2}}    ~.\ee 
In the following we consider the limit where $\alpha$ is small compared with 1, in which case $\alpha \approx B^{2}$. 

             For the case of F-term inflation with suppressed order $H$ corrections to the A-terms, fluctuations $\delta \theta$ of the phase of the flat-direction field are unsuppressed. At the onset of coherent oscillations, the A-term induce an amplitude fluctuation $\delta N/N = O(1) \delta \theta$ \cite{pcurv}. The fluctuation of the phase is related to the fluctuation of the field during inflation by $\delta \theta \approx \delta N_{2}/N_{I}$, where $N_{2}/\sqrt{2}$ is the complex part of the RH sneutrino field. Therefore, up to an O(1) factor, the amplitude fluctuation of the RH sneutrino at the onset of coherent oscillations is the same as in D-term inflation.

            The value of $N_{I}$ in D-term inflation has an upper limit from the requirement that 
$V^{''}(N) \lae H^{2}$, above which the RH sneutrino is no longer effectively massless and is rapidly evolving. (In this we are discounting the unlikely possibility that the RH sneutrino is rapidly rolling at $N \approx 60$ e-foldings before the end of inflation.) This corresponds to $N_{I} \lae N_{*}$ where
\be{e27}  N_{*} = \left(\frac{48}{5 \lambda_{N}^{2}} \right)^{1/4}
 \left(M H_{I}\right)^{1/2}      ~.\ee 
$N < N_{*}$ then implies a lower bound, $\alpha_{min}$, on the value of $\alpha$, given by  
\be{e28}  \alpha_{min} \approx \left(\frac{5}{48}\right)^{1/2}  \frac{\lambda_{N}}{\pi^{2} P_{{\cal R}}} \frac{H_{I}}{M}   ~.\ee
The 2-$\sigma$ observational upper bound for an uncorrelated CDM isocurvature perturbation from WMAP3 is $\alpha_{lim} = 0.26$ \cite{bean} \footnote{An alternative method, using Bayesian model selection, may allow a tighter upper bound on $\alpha$ to be obtained \cite{trotta}.}.  Requiring that 
$\alpha_{min} < \alpha_{lim}$ then gives us an upper bound on the value of $H$ during inflation 
\be{e29} H_{I} \lae \left(\frac{48}{5}\right)^{1/2} \frac{\pi^{2} P_{{\cal R}} M \alpha_{lim}}{\lambda_{N}}  \equiv 4.4 \times 10^{11} 
 \left(\frac{0.1}{\lambda_{N}}\right) \left(\frac{\alpha_{lim}}{0.26}\right) \GeV   ~.\ee 

      F-term inflation differs from D-term inflation in that $N_{I}$ is fixed to be at the minimum of its potential during inflation, 
\be{e31}  N_{I} = N_{min} \equiv  \left(\frac{48 c_{N}}{\lambda_{N}^{2}} \right)^{1/4}
 \left(M H_{I} \right)^{1/2}      ~.\ee 
The lower bound on $\alpha$ and the upper bound on $H_{I}$ in this case are now 
\be{e32}  \alpha_{min} = \left(\frac{1}{48 c_{N}}\right)^{1/2} 
\frac{\lambda_{N}}{\pi^{2} P_{{\cal R}} } \frac{H_{I}}{M}   ~\ee
and
\be{e33} H_{I} \lae \left(24 c_{N}\right)^{1/2} \frac{\pi^{2} P_{{\cal R}} \alpha_{lim} M}{\lambda_{N}} \equiv 9.5 \times 10^{11} c_{N}^{1/2} 
\left(\frac{0.1}{\lambda_{N}}\right)  \left(\frac{\alpha_{lim}}{0.26}\right) \GeV   ~.\ee 

      It is interesting that the upper bounds on $H_{I}$ are typically in the range $10^{10} - 10^{13} \GeV$ for $\lambda_{N} = 0.01 - 1$. As such they are large enough not to require an inflation model with an extremely small value of $H_{I}$, yet small enough to offer a realistic prospect of detecting the CDM isocurvature perturbation in the future.          

      We can compare these upper bounds with the value of $H_{I}$ expected in common SUSY inflation models. In the case of D-term hybrid inflation \cite{dti}, $H_{I} \approx 1 \times 10^{13}g \GeV$, where $g$ is the $U(1)$ gauge coupling, corresponding to an energy density during inflation $V = g^{2} \xi^{2}/2$. (In this we have assumed a Fayet-Iliopoulos term $\xi^{1/2} = 8 \times 10^{15} \GeV$, as required by the observed curvature perturbation.) Comparing with \eq{e29} we see that for D-term hybrid inflation the CDM isocurvature perturbation is compatible with observation only if the gauge coupling satisfies $g \lae 0.05$ when $\lambda_{N} \approx 0.1$. This suggests that the CDM isocurvature perturbation in D-term hybrid inflation is likely to be close to the present observational limit, although a small value of the gauge coupling and/or non-renormalisable superpotential coupling is required for the CDM isocurvature perturbation to be within observational limits; if $\lambda_{N} \approx 1$ and $g \gae 0.1$ then D-term hybrid inflation would be excluded. In the case of F-term hybrid inflation with suppressed order $H$ A-term corrections, $|S|/M \lae 1/40$ is necessary in order to have a large enough correlation length for an isocurvature fluctuation to exist. The value of $S$ at $N \approx 60$ e-foldings of inflation is $|S|/M = \kappa \sqrt{N}/2\pi \approx 1.2 \kappa$, where $\kappa$ is the Yukawa coupling of the F-term hybrid inflation superpotential ($W = \kappa S(-\mu^{2} + \overline{\phi} \phi)$) \cite{fti}. Therefore to have  $|S|/M \lae 1/40$ we require that $\kappa \lae 0.01$. The value of $H_{I}$ during F-term hybrid inflation is $H_{I} = 7 \times 10^{12}  \kappa \GeV$, corresponding to an energy density during inflation $V = \kappa^{2} \mu^{4}$. (We have assumed $\mu = 5 \times 10^{15} \GeV$ as required by the observed curvature perturbation.) Therefore if a CDM isocurvature perturbation exists ($\kappa \lae 0.01$) then $H_{I} \lae 7 \times 10^{10} \GeV$. Comparing with \eq{e33} shows that the CDM isocurvature perturbation in F-term hybrid inflation can be close to the present observational limit if $\lambda_{N} \approx 1$.

          In the above we have assumed that there is no suppression of the isocurvature perturbation after inflation. 
This will generally be true in the case without order $H$ corrections to the A-term. In this case the isocurvature perturbation originates from phase fluctuations of the flat direction field, which do not evolve since the phase field is massless. In the case of D-term inflation with order $H$ corrections to the A-terms after inflation we need to check whether there is any suppression of the amplitude and phase fluctuations. 

            Once inflation ends the potential acquires a minimum to which the flat direction field will roll and coherently oscillate. We consider the evolution of the RH sneutrino along the real direction,  
$ N(\vec{x},t) = N(t) + \delta N(\vec{x},t)$. The zero mode, $N(t)$, and $k$ mode of $\delta N(\vec{x},t)$, $\delta N_{k}(t)$, will satisfy the equations  
\be{e35}   \ddot{N} + 3 H \dot{N} = -\frac{\partial V}{\partial N}     ~\ee 
and
\be{e36}   \delta \ddot{N}_{k} + 3 H \delta \dot{N}_{k}  \approx   - \left(\frac{\partial^{2} V}{\partial N^{2}}\right)_{N(t)}  \delta N_{k}    ~,\ee
where we assume that $k/a \ll H$ in \eq{e36}. 
If at the end of inflation $N(t) \ll N_{min}(t)$ then $N(t)$ and $\delta N_{k}(t)$ will evolve in a potential 
$V \approx - c_{N}H^{2}|N|^{2}$. As a result $N(t)$ and $\delta N_{k}(t)$ satisfy the same equation and therefore $\delta N_{k}(t)/N(t)$ remains constant until $N(t) \approx N_{min}$. Thereafter, to a good approximation, $N(t) = N_{min}(t)$ during the post-inflation era. In this case   
\be{e37}   \delta \ddot{N}_{k} + 3 H \delta \dot{N}_{k} = -4 c_{N} H^{2} \delta N_{k}    ~.\ee 
With $H \propto a^{-3/2}$ during inflaton coherent oscillations the solution is $\delta N_{k}(t) \propto a^{\gamma}$, where \cite{drt,rhsnc}  
\be{e38}   \gamma  =  \frac{1}{2} \left[ -\frac{3}{2} + \sqrt{ \frac{9}{4} - 16 c_{N}} \right]    ~.\ee 
The largest suppression will occur for the case where $16 c_{N} > \frac{9}{4}$, in which case  
$|\delta N_{k}| \propto a^{-3/4}$. For a $d = 4$ flat direction $N(t) \approx N_{min}(t) \propto H^{1/2} \propto a^{-3/4}$, 
therefore $\delta N_{k}/N_{min}$ is constant i.e. there is no suppression of the isocurvature perturbation for a $d = 4$ flat direction due to evolution of the fields after inflation. Note that this result is peculiar to a $d = 4$ flat direction: in general the 
minimum of the potential is proportional to $H^{1/(d-2)}$, so that $\delta N_{k} / N_{min} \propto a^{3(1/(d-2)-1/2)/2}$. 

   In the case where there are order $H$ corrections to the A-terms after inflation, the phase field at the minimum of the potential will also have a mass squared of order $H^{2}$ and so will oscillate about the minimum is the same way as the amplitude field. Therefore there will also be no suppression of the magnitude of the phase fluctuation after inflation.    
 
              To summarise, a CDM isocurvature perturbation will be generated in the case of D-term inflation models in general and also in F-term inflation model with suppressed order $H$ corrrections to the A-terms. The magnitude of the CDM isocurvature perturbation
in common SUSY inflation models can be naturally close to the present observational bound. 

\subsection{Baryon Isocurvature Perturbation} 

          Fluctuations of the phase and amplitude of the AD scalar will result in baryon isocurvature perturbations. We will consider only the case of phase fluctuations, which will be unsuppressed if order $H$ corrections to the A-terms are suppressed during inflation. The magnitude of the baryon isocurvature perturbation due to unsuppressed amplitude fluctuations in D-term inflation models will be similar. 

            In the case of D-term inflation, there is no suppression of fluctuations in the real or imaginary direction during inflation. Defining the AD scalar by 
\be{e39a} \Phi =  \frac{\phi}{\sqrt{2}}e^{i \theta} = \frac{1}{\sqrt{2}} \left(\phi_{1} + i \phi_{2}\right)   ~,\ee
the fluctuation of the phase is related to the field fluctuations by  
\be{e39}  \delta \theta =  \frac{- \sin \theta \delta \phi_{1} + \cos \theta \delta \phi_{2}}{\phi_{I}}       ~, \ee
where $\phi_{I}$ is the value of $\phi$ during inflation.   
Therefore
\be{e40} <\delta \theta ^{2}> =   \frac{\sin^{2} \theta <\delta \phi_{1}^{2}> + \cos^{2} \theta <\delta \phi_{2}^{2}>}{\phi_{I}^{2}}       ~,\ee 
where $\delta \phi_{1}$ and $\delta \phi_{2}$  are uncorrelated perturbations.  Using $<\delta \phi_{1}^{2}> 
= <\delta \phi_{2}^{2}> = <\delta \phi^{2}>$ we obtain 
\be{e41} <\delta \theta ^{2}> =   \frac{<\delta \phi^{2}>}{\phi_{I}^{2}}       ~.\ee
The power spectrum of $\theta$ fluctuations is therefore  
\be{e42} P_{\delta \theta} =  \frac{P_{\delta \phi}}{\phi_{I}^{2}} \;\; ; \;\;\; 
P_{\delta \phi} = \left(\frac{H_{I}}{2 \pi} \right)^{2}     ~.\ee 
The Fourier transform of the $\delta \theta$ power spectrum is 
\be{e43}  |{\cal T}_{\theta}|^{2} =  \frac{2 \pi^{2}}{k^{3}} \frac{P_{\delta \phi}}{\phi_{I}^{2}}  =    
\frac{1}{2 k^{3}} \frac{H_{I}^{2}}{\phi_{I}^{2}}     ~.\ee 
From \eq{e10} the baryon asymmetry is proportional to $\sin (2\theta)$. Therefore 
\be{e44}  \frac{\delta n_{B}}{n_{B}} = f_{\theta} \delta \theta \;\;\ ; \;\;  f_{\theta} = \frac{2}{\tan 2 \theta}    ~.\ee  
The baryon entropy perturbation is ${\cal S}_{B} = \delta n_{B}/n_{B}$. 
The Fourier transform of the baryon entropy perturbation power spectrum is then
\be{e45} |{\cal S}_{B}|^{2} = f_{\theta}^{2} |{\cal T}_{\theta}|^{2} 
= \frac{1}{2 k^{3}} \frac{f_{\theta}^{2} H_{I}^{2}}{\phi_{I}^{2}}        ~.\ee
Therefore the contribution of the baryon isocurvature (BI) perturbation to the CMB is given by 
\be{e46} \alpha_{BI} \approx  \left(\frac{\Omega_{B}}{\Omega_{DM}}\right)^{2} \frac{|{\cal S}_{B}|^{2}}{|{\cal R}|^{2}} = \left(\frac{\Omega_{B}}{\Omega_{DM}}\right)^{2}  \frac{f_{\theta}^{2} H_{I}^{2}}{4 \pi^{2} P_{{\cal R}} \phi_{I}^{2}}    ~,\ee
where the factor $(\Omega_{B}/\Omega_{DM})^{2}$ rescales the CDM isocurvature perturbation $C_{l}^{iso}$ to the baryon case \cite{crotty}.   

                 As in the case of the RH sneutrino, in the case of D-term inflation there is an upper limit on $\phi_{I}$ from the requirement that $\phi_{I} \lae \phi_{*}$, defined by $V^{''}(\phi_{*}) = H^{2}$, 
\be{e47}   \phi_{*}^{2} = \left(\frac{48}{5 \lambda_{\Phi}^{2}} \right)^{1/2} H_{I} M    ~.\ee 
$\alpha_{BI}$ can then be written as  
\be{e48}    \alpha_{BI} \approx 
 \left(\frac{\Omega_{B}}{\Omega_{DM}}\right)^{2}
\left( \frac{5 \lambda_{\Phi}^{2}}{48} \right)^{1/2} 
  \frac{ f_{\theta}^{2} }{4 \pi^{2}  P_{{\cal R}} } \frac{H_{I}}{M} 
\left( \frac{\phi_{*}}{\phi_{I}} \right)^{2}        ~.\ee 
The same observational limit, $\alpha_{lim} < 0.26$, applies to the baryon isocurvature perturbation as in the CDM case, since the perturbations are almost observationally indistinguishable  \cite{bean}. 
This then gives an upper bound on $H$ during inflation 
\be{e49}   H_{I} \lae 5.7 \times 10^{12} \left( \frac{0.04}{\Omega_{B}} \right)^{2} \left( \frac{\Omega_{DM}}{0.23} \right)^{2} 
\frac{1}{ \lambda_{\Phi} f_{\theta}^{2}}  \left( \frac{\alpha_{lim}}{0.26} \right) \left( \frac{\phi_{I}}{\phi_{*}} \right)^{2}  \GeV ~.\ee
Therefore in the case where $\phi_{I}$ is close to $\phi_{*}$, and with $\lambda_{\Phi} \lae \lambda_{N}$ as required by the baryon-to-dark matter ratio, the constraint on $H_{I}$ is much weaker than in the case of the CDM isocurvature perturbation, \eq{e29}. However, in D-term inflation it is possible that $\phi_{I}$ could be small compared with $\phi_{*}$ while $N_{I}$ is close to $N_{*}$, in which case it may be possible to have both a baryon and CDM isocurvature perturbation of similar magnitude.        

                 In the case of F-term inflation we should replace $\phi_{I}$ with $\phi_{min}$ in \eq{e49}, which gives 
\be{e50} H_{I} \lae  1.3 \times 10^{13} \left( \frac{0.04}{\Omega_{B}} \right)^{2} \left( \frac{\Omega_{DM}}{0.23} \right)^{2} 
\frac{c_{\phi}^{1/2}}{ \lambda_{\Phi} f_{\theta}^{2}}  \left( \frac{\alpha_{lim}}{0.26} \right) \GeV ~.\ee 
This is generally a much weaker bound than the CDM isocurvature upper bound, \eq{e33}, when $\lambda_{\Phi} \lae \lambda_{N}$. Thus the baryon isocurvature perturbation will make a much smaller contribution than the CDM isocurvature perturbation in F-term inflation models.  The observation of CDM and baryon isocurvature perturbations of a similar magnitude would therefore be a signature of inflation driven by a D-term in this model.

\section{Conclusions}

             In trying to reconstruct early cosmology from information provided by astronomical observations and particle physics experiments, we need to make full use of any clue provided by nature. One possible clue is the striking similarity of the density in baryons and cold dark matter. It is a highly non-trivial feature of a particle physics theory to have a mechanism within its structure which can explain the baryon-to-dark matter ratio. Therefore the baryon-to-dark matter ratio could serve a discrimiator of, simultaneously,  the correct particle physics theory, the nature of cold dark matter and the origin of the baryon asymmetry. Moreover, if we accept that a model which is able to account naturally for the baryon-to-dark matter ratio has a high likelihood of being the correct model, we may be able to use this information to draw conclusions about the very early Universe, such as the nature of the inflation model.  

           In this paper we have considered the implications of the baryon-to-dark matter ratio in the context of the $\nu$MSSM. The RH sneutrino condensate appears to be a uniquely suitable candidate for cold dark matter in the $\nu$MSSM 
from the point of view of the baryon-to-dark matter ratio. In the case where both the baryon asymmetry and CDM density originate along $d = 4$ flat directions of the $\nu$MSSM, we have shown that the observed baryon-to-dark matter 
ratio can be understood if there is a small hierarchy between the non-renormalisable couplings of the RH neutrino and $(H_{u}L)^{2}$ flat direction superfields, with the $(H_{u}L)^{2}$ coupling one or two orders of magnitude less than the RH neutrino coupling. Such a hierarchy of couplings seems plausible given the range of values observed for the renormalisable Yukawa couplings of the MSSM. It would be significant if this hierarchy could be understood from the point of view of a complete theory of Planck-scale physics such as string theory.   

           Observation of a CDM and/or baryon isocurvature perturbation, combined with indirect accelerator evidence for a very weakly coupled RH sneutrino LSP, such as a coloured or charged MSSM-LSP \cite{moroi}, would provide strong support for the RH sneutrino condensate dark matter model. The model indicates that the CDM isocurvature perturbation in D-term inflation models and in F-term inflation models with sufficiently suppressed order $H$ A-term corrections is likely to be close to present observational limit. In addition, it is possible that a large baryon isocurvature perturbation could arise in D-term inflation models, but in F-term inflation models it is generally suppressed relative to the CDM isocurvature perturbation. Therefore observation of a CDM and baryon isocurvature perturbation of similar magnitude would be a signature of D-term inflation in this context, thus providing us with direct information on the nature of SUSY inflation. 

\subsection*{Acknowledgement} 

     The author would like to thank Christoph Luhn for his comments.

\end{document}